\documentclass{elsart}

\usepackage{natbib}

\usepackage{graphics}

\usepackage{amssymb}

\begin{document}

\begin{frontmatter}



\title{A way to synchronize models with seismic faults for earthquake
forecasting: Insights from a simple stochastic model}

\author[address1]{\'Alvaro Gonz\'alez\corauthref{Alvaro}}
\ead{Alvaro.Gonzalez@unizar.es}
\thanks[Alvaro]{Tel. +34 610329045; fax. +34 976761106}
\author[address2]{Miguel V\'azquez-Prada}
\author[address1]{Javier B. G\'omez}
\author[address2]{Amalio F. Pacheco}
\ead{amalio@unizar.es}
\address[address1]{Departamento de Ciencias de la Tierra. Universidad de
  Zaragoza. C./ Pedro Cerbuna, 12. 50009 Zaragoza, Spain.}
\address[address2]{Departamento de F\'isica Te\'orica and BIFI. Universidad de
  Zaragoza. C./ Pedro Cerbuna, 12. 50009 Zaragoza, Spain.}

\begin{abstract}
Numerical models are starting to be used for determining the
future behaviour of seismic faults and fault networks. Their final
goal would be to forecast future large earthquakes. In order to
use them for this task, it is necessary to synchronize each model
with the current status of the actual fault or fault network it
simulates (just as, for example, meteorologists synchronize their
models with the atmosphere by incorporating current atmospheric
data in them). However, lithospheric dynamics is largely
unobservable: important parameters cannot (or can rarely) be
measured in Nature. Earthquakes, though, provide indirect but
measurable clues of the stress and strain status in the
lithosphere, which should be helpful for the synchronization of
the models.

The rupture area is one of the measurable parameters of
earthquakes. Here we explore how it can be used to at least
synchronize fault models between themselves and forecast synthetic
earthquakes. Our purpose here is to forecast synthetic earthquakes
in a simple but stochastic (random) fault model. By imposing the
rupture area of the synthetic earthquakes of this model on other
models, the latter become partially synchronized with the first
one. We use these partially synchronized models to successfully
forecast most of the largest earthquakes generated by the first
model. This forecasting strategy outperforms others that only take
into account the earthquake series. Our results suggest that
probably a good way to synchronize more detailed models with real
faults is to force them to reproduce the sequence of previous
earthquake ruptures on the faults. This hypothesis could be tested
in the future with more detailed models and actual seismic data.
\end{abstract}


\end{frontmatter}





\section{Introduction: Data assimilation in dynamical fault models}

Numerical models are now frequently used to simulate the seismic
behaviour of faults \citep[e.g.][]{Kato601,Fitzenz602,Kuroki598}
and fault networks
\citep[e.g.][]{Ward542,Hashimoto597,RobinsonBenites,Rundle593,Soloviev2003,Robinson608,Rundle603}.
In these models, fault planes separate lithospheric blocks that
are strained at specific rates, and sudden slips (earthquakes) are
generated by the faults according to certain friction and/or
rupture laws. Although no completely realistic dynamical model
presently exists, these simulations are now sufficiently credible
to begin to play a substantial role in scientific studies of
earthquake probability and hazard \citep{Ward542}. The final goals
of the numerical modelling of seismicity are not different from,
for example, the goals of numerical models of the atmosphere. A
good model should be able to:
\begin{enumerate}
\item reproduce the general characteristics of the system, \item
mimic the state of the system at the present moment, and \item
forecast the future evolution of the system.
\end{enumerate}
Most numerical models of seismicity have been designed to achieve
the first goal, by reproducing general characteristics of
earthquakes such as their size-frequency distribution
\citep[e.g.][]{Bak,Olami592,Dahmen9,Preston383,VazquezPrada250},
or the generation of aftershock and foreshocks
\citep[e.g.][]{Hainzl599}. When a model is designed this way, it
is left to evolve freely according to its rules, and all that is
checked is whether the overall results of the model are similar to
the observations made in Nature or not.

The second goal requires data assimilation, that is, the process
of absorbing and incorporating observed information into the
model. By this process, the model is tuned and synchronized, at
least partially, with the real system it tries to simulate. In a
meteorological model, data of atmospheric pressure, temperature,
humidity, cloud cover, precipitation, etc. measured in a given
moment at different locations and heights can be included. With
this procedure, the model becomes a reasonably good representation
of the atmosphere at that moment. Then it can be used to calculate
the probable future atmospheric evolution (i.e. the third goal
cited above).

Seismic data assimilation poses greater problems than its
meteorological equivalent. This explains (at least partially) the
relative delay in developing reliable forecasts of large
earthquakes. The inner workings of both the atmosphere
\citep{Houghton} and the lithosphere
\citep{Goltz,Turcotte,KeilisBorok90} are complex and chaotic, so
they are inherently difficult to forecast. However, while
meteorologists can probe the atmosphere every day at different
places and heights (and assimilate the obtained data in their models
in near real-time), lithospheric variables of paramount importance,
such as the stress and strain, can be measured only in certain
places, and not at any time: earthquakes have unobservable dynamics
\citep{Rundle}. For example, the best current compendium of stress
magnitudes and directions in the lithosphere is the World Stress Map
\citep{Zoback1992,Reinecker594}, whose entries are point static
time-averaged estimates of maximum and minimum principal stresses in
space. And the direct measurements of stress on active fault zones
at depth are still scarce
\citep[e.g.][]{Ikeda2001,Tsukahara2001,Yamamoto2001,Hickman591,Boness604}.
The dynamical models would need better spatial and temporal
information of stress, both more abundant and more systematically
collected than that currently available \citep{Rundle603}. It is
thus necessary to seek ways to tune and synchronize the models with
more abundant observable data.

A first step of data assimilation in models of earthquake faults
is to introduce information regarding the topology (that is, the
shape and location) of the active faults and their long-term
behaviour. For example, the long-term fault slip rate, and the
average recurrence interval of the largest earthquakes in the
fault can be estimated from paleoseismological studies and should
be included in the models \citep{Grant2004}. Examples of this
approach are the works of \cite{Rundle593,Rundle603} and
\cite{Robinson608}. The surface deformation measured via Global
Positioning System (GPS) networks and by Synthetic Aperture Radar
Interferometry (InSAR) can also constitute input data for the
dynamical fault models \citep{Rundle603}. Earthquakes themselves
are indeed the most obvious observable events of lithospheric
dynamics, and could provide the most detailed data available to
assimilate in the models, but how? The earthquake rupture area
could be an important clue.

The rupture area and slip distribution in real earthquakes can be
very complex \citep{Sieh258,Kanamori606}, but can be estimated in
a variety of ways. The actual slip distribution can be obtained by
inverting the observed seismic waveforms \citep{Kanamori606} or
tsunami waveforms \citep[e.g.][]{Tanioka,Baba}, and/or by geodetic
modelling of surface displacement \citep{Yabuki}. Some earthquakes
produce surface ruptures, which are useful for estimating the
rupture area \citep{Stirling609}. Although most surface ruptures
occur in large shocks, with magnitudes larger than about 6, they
have been reported for earthquakes with magnitudes down to 2.5
\cite[see the compilation of historic earthquakes with surface
rupture by][pp. 473-485]{Yeats}. Also, the rupture area can be
estimated from the seismic moment \citep[calculated from the
amplitude spectra of seismic waves;][]{Scholz,Kanamori606}, or
from the moment magnitude \citep{Wells213,Stirling609,Dowrick607}.
Frequently the location of early aftershocks is used to determine
the rupture area of the mainshock \citep{Wells213}, although the
aftershock zone tends to grow with time \citep{Kisslinger} and is
not necessarily a good indicator of that area \citep{Yagi1999}.

Complex models with realistic fault topology are able to reproduce
the rupture area and coseismic slip of historical earthquakes. It
is thus possible to force the model to reproduce the rupture of a
historical earthquake, and let it evolve from that moment onwards
to see what could happen in the future. For example,
\cite{Ward542} developed a model including the network of main
faults in the San Francisco Bay Area (California). He forced the
model to reproduce the San Andreas Fault surface coseismic slip of
the 1906 San Francisco earthquake, and let it evolve freely from
that earthquake onwards, in an attempt to simulate the probable
sequence of earthquake ruptures during the next 3000 years.

But considering only the data of the largest earthquake in the
series is probably not sufficient to properly synchronize the
model. Complex and chaotic systems are very sensitive to the
initial conditions. The information regarding only one event
probably does not sufficiently constrain the initial conditions,
and the calculated evolution will probably be a particular case of
a large range of possible outcomes. Will this panorama improve by
forcing the model to reproduce all the observed earthquake
ruptures, including the small ones? Probably yes. To check whether
this idea works, at least to forecast synthetic seismicity, is the
purpose of this paper. The number of recorded large earthquakes is
relatively scarce, especially in individual faults (where the
recorded series very rarely includes ten large events). This
hampers the ability to characterize statistically the
effectiveness of any forecasting method. Synthetic earthquake
catalogues, on the other hand, can be as long as desired. This
enables to ascertain, with robust statistics, whether a
forecasting strategy could be useful, before endeavouring to apply
it to real seismicity.

In the following sections, our goal will be to forecast the
largest earthquakes generated by the minimalist model, a simple
numerical fault model. We will show that when all the earthquake
ruptures generated by this model are imposed on other, similar
models, these become partially synchronized with the former. We
use them to declare alarms that efficiently mark the occurrence of
the largest shocks in the first model. The results are much better
than those obtained with other strategies that consider only the
earthquake series. The model, albeit simple, is stochastic (it
involves randomness), so its efficient forecasting is not trivial.
We will describe how this stochasticity can be dealt with, by
using an approach similar to the so-called ensemble forecasting
used in Meteorology \citep{Palmer615}. The method could be used in
other more detailed and realistic models (stochastic or not) to
test our general conclusion: that they might be partially
synchronized with actual faults by being forced to reproduce the
series of observed earthquake ruptures.

In the next section we describe the model and its properties.
Then, we outline the general scheme of prediction and the
forecasting strategies used as reference to assess the merits of
any other predictive method in the model. Finally, the method
based on partial synchronization is explained and its possible
utility discussed.

\section{The minimalist model}

The minimalist model is the numerical model whose largest
earthquakes we will try to forecast. It was introduced in a
previous work \citep{VazquezPrada250}, and has mainly two,
apparently contradictory, advantages for the purpose of this
paper: it is simple but, at the same time, it is difficult to
forecast. Because it is simple, several of its properties can be
derived analytically, and it can be characterized in detail with
numerical simulations which do not require an impractical amount
of computer time. Because it is stochastic, it is difficult to
forecast, so the results we will obtain here are not trivial. In
the following paragraphs we will explain how the model works, and
what are its main properties, comparing them with those of actual
faults.

\subsection{How the model works}
The model is a simple (hence its name) cellular automaton.
Cellular automata are frequently used to model seismic faults. In
these models, the fault plane is divided into a grid of cells
(each cell representing a fraction of the fault area), and the
time evolves in discrete time steps. Each cell's state is updated
at each time step according to rules that usually depend on the
state of the cell or that of its neighbors in the previous time
step. These rules can be designed according to certain friction
laws \citep{BenZion493}, stress transfer
\citep{Olami592,Hainzl599,Preston383}, and the effect of fluids
\citep{Miller355}. In the minimalist model, as well as in other
very simple cellular automata \citep[e.g.][]{Newman478,Box}, these
details are ignored: the model is driven stochastically, there are
only two possible states for each cell, and the earthquakes are
generated according to simplified breaking rules.

Let us now explain the simplified view of earthquake generation
that the model tries to sketch. In actual faults, the regional
stress strains the rock blocks of the fault, making portions
(patches) of the fault plane to become metastable. That is, they
are static, but store enough elastic energy to propagate an
earthquake rupture once triggered. Different processes (for
example, fault creep --aseismic slip-- and plastic deformation)
dissipate stress along the fault plane, so stress is not directly
converted into elastic strain. Earthquakes rupture some of the
metastable patches of the fault, that then become stable, thus
relieving strain. The hypocentre of an earthquake is usually
located in a particularly strong patch of the fault plane, called
``asperity'' \citep{KanamoriStewart,Aki610,Das365,AsperityLei}.
Asperities appear to be persistent features where earthquake
ruptures start once and again \citep{Aki610,Okada611}. Once the
rupture starts, it propagates along the fault plane until it
arrives at a patch of the fault that is not sufficiently strained.
Then the rupture cannot propagate further, and is arrested. The
relatively stable patch that is not sufficiently strained and that
arrests the rupture is called the ``barrier''
\citep{DasAki,Aki610,Das365}.

The model, depicted in Fig. 1, sketches these features as follows.
It divides the plane of a fault into an array of $N$ equal cells,
each denoted by an index $i$. In previous papers
\citep{VazquezPrada250,MM2,MM3,Gomez546,Europhysics}, this array
was drawn vertically, in order to simplify its mathematical
description. Here the model will be drawn horizontally, in order
to sketch the fault plane in a way more similar to that of actual
faults (which are usually longer along the strike than along the
dip). Some other cellular automaton models discretize the fault
plane in a similar way \citep[e.g.][]{Rundle603}. The parameter
$N$ is the only one that can be changed in the model. The cells
can only be in one of two states: ``empty'' (stable) or
``occupied'' (metastable). The state of the model at each time
step can be described simply by stating which cells are occupied
and which are not. The increase of regional stress, as in other
simple models \citep{Bak,Newman478,Castellaro441,Box}, is
represented by the random addition of ``stress particles''. This
randomness is a way of dealing with the complex stress increase in
actual faults. At each time step, one cell is selected randomly,
and a new particle arrives on it. That is, each cell has the same
probability, $1/N$ of receiving the new stress particle. If the
chosen cell is empty, the particle ``occupies'' it. This means
that the regional stress has produced enough strain on that cell
to make it metastable. If the cell is already occupied, that
stress particle is lost; this is analogous to stress dissipation
on the fault plane. The total number of occupied cells represents
the total elastic strain on the fault.

\begin{figure}
\begin{center}
\includegraphics{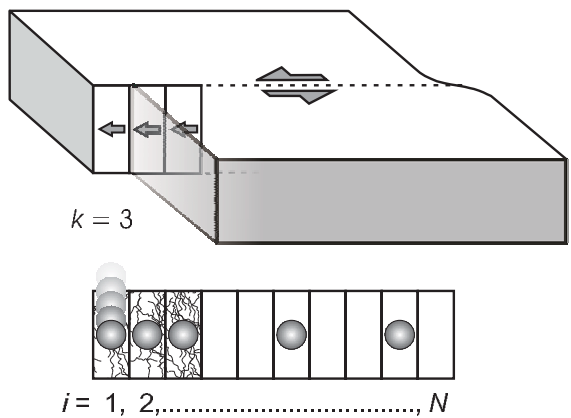}
\end{center}
\caption{The minimalist model as a sketch of a seismic fault. The
fault plane is divided into an array of $N$ equal cells, denoted
with an index $i$, $1\le i \le N$. The increase of regional stress
is represented by the random addition of ``stress particles'' to
the cells. Earthquake ruptures start at an asperity, the cell
$i=1$, when a stress particle arrives to it. The rupture
propagates through all the consecutive metastable cells (occupied
by particles). The rupture area is $k$, the number of cells
broken. The figure depicts an earthquake with $k=3$.}
\end{figure}

In the model, we assume that there is only one, persistent,
asperity: the first cell, $i=1$, placed at one end of the array.
This option is chosen because it simplifies the analytical
description of the model. When a stress particle fills cell $i=1$,
a rupture starts there, and propagates through all the consecutive
metastable cells until it is arrested by a stable cell. That is,
if all the successive cells $i=1$ to $i=k$ are occupied, and cell
$k+1$ is empty, then the effect of the earthquake is to empty all
the cells from $i=1$ to $i=k$. The other cells, $i>k$ remain
unaltered. The cell $k+1$ is a barrier: it is empty (stable), so
the rupture cannot propagate through it. The size (rupture area)
of the earthquake is $k$, the number of cells broken in the
synthetic earthquake. Thus, the earthquake size in the model is
discrete, $1\leq k \leq N$. Earthquakes, in practice, are
instantaneous in the model (they do not last for any time step).
This represents the fact that earthquake ruptures are, indeed,
much faster than the slow stress loading represented by the
addition of particles.

The random addition of particles is what makes the model
stochastic. It also determines the rate at which earthquakes occur
in the model. At each time step, independently of the previous
earthquake history, there is a probability $1/N$ for the incoming
stress particle to arrive at cell $i=1$ and start an earthquake.
Thus an earthquake, on average, occurs every $N$ steps. The time
between any two consecutive earthquakes is purely random
(Poissonian, with rate $1/N$).

The cellular-automaton approach of this model is similar to that
of the ``forest fire'' models, in which clusters of interconnected
occupied cells (``trees'') ``burn'' and are reset to empty when
they are randomly struck by ``lightning''
\citep{Drossel407,Henley417}. The utility of this kind of models
for earthquake physics has been noted by \cite{Rundle}. In the
minimalist model there is no random ``lightning'': the clusters of
interconnected metastable sites are only emptied if they are
connected to the cell $i=1$ \textit{and} if this fails.

\subsection{Main properties of the model}\label{sec:quiescence}

The minimalist model, because of its extreme simplicity, lacks the
detailed description of the seismic process that a fully dynamical
model can display. For example, it does not include the effects of
fault friction, elastic stress transfer, or the role of fluids
that more complex models can take into account. However, it
spontaneously displays several properties that are comparable to
those of actual faults, outlined as follows:

\begin{enumerate}

\item {\it Earthquake size-frequency distribution}. It is of the
characteristic-earthquake type
\citep{Wesnousky422,SC1984,Youngs308,Wesnousky44}, observed in
seismic faults with simple traces  \citep{Stirling} and in other
numerical models if the fault plane is homogeneous
\citep[e.g.][]{Rundle1993,Main296,Dahmen9,Steacy,Moreno2,Hainzl613,Heimpel351,Zoller614}.
In this distribution there is a relative excess of events (called
characteristic earthquakes) which break the whole fault or most of
it. In the model, they are the earthquakes with size $N$, and will
be the events to forecast. The Gutenberg-Richter distribution
\citep{Ishimoto,Gutenberg1,Gutenberg2} observed in regional
seismicity (which includes contributions from many faults) can be
reproduced adding up the seismicity of an ensemble of minimalist
models whose sizes ($N$) are distributed as in actual faults
\citep{MM3}.

\item {\it Duration of the earthquake cycle}. The earthquake cycle
of a fault is the time interval between two consecutive
characteristic earthquakes \citep[e.g.][]{Scholz311}. The
statistical distribution of these intervals in the model is
similar to the observed in seismic faults \citep{Gomez546}. The
distribution of time intervals between consecutive earthquakes of
any size in the model is Poissonian distributed. However, if only
the characteristic earthquakes are considered, the distribution is
not Poissonian. This is because the maximum possible size of an
event depends on the size of the previous event and the time
elapsed since it occurred. This is commented in the next
paragraph.

\item {\it Stress shadow}. When a fault generates a large
earthquake, the elastic strain is reduced, and a minimum time has
to elapse until the fault, by slow tectonic deformation,
accumulates enough strain to generate another large earthquake.
This effect is called stress shadow \citep{Harris366}. In the
minimalist model there is a stress shadow: if an event of size $k$
takes place, at least $k$ time steps have to elapse until another
event of that size can occur.

\item {\it Pattern of strain loading}. In actual faults, the
strain increases rapidly just after a large earthquake, and then
more slowly \citep{Michael2005}. In the model, the total elastic
strain is represented by the occupation (the total number of
occupied cells, Fig. 2), which has a similar pattern. Just after a
large earthquake, there are fewer occupied cells, so it is more
probable for the incoming particles to land on empty cells, and
the occupation grows faster than later on (Fig. 2).

\begin{figure}
\begin{center}
\includegraphics{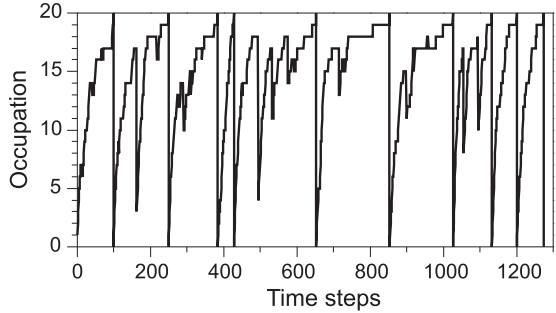}
\end{center}
\caption{The number of occupied cells in a minimalist model with
$N=20$, for ten seismic cycles. This number is analogous to the
total elastic strain accumulated in the fault. Sudden drops
correspond to earthquakes. Each seismic cycle ends with an
earthquake of size $N$.}
\end{figure}

\item {\it Seismic quiescence}. The model displays seismic
quiescence (absence of earthquakes) before the characteristic
events. Once $N-1$ cells (from $i=2$ to $i=N$) become occupied,
the occupation reaches a plateau (Fig. 2) and, on average, $N$
time steps have to elapse until the next earthquake (which is the
characteristic one) occurs. Seismic quiescence has been observed
preceding many large earthquakes \citep[e.g.][]{Wyss1988,Scholz},
although in other cases the opposite effect (increased activity)
has been observed \citep[e.g.][]{Bowman,Reasenberg,Tiampo}. The
minimalist model does not show this last behaviour.
\end{enumerate}

\section{General scheme of forecasting}\label{sec:forecasting}

In this section we will explain the general framework for the
forecasting of the largest earthquakes in the model. As a first
remark, we have to consider that the model is stochastic, so it is
not predictable with absolute precision. Only simple deterministic
systems are fully predictable. The evolution of complex systems,
such as the atmosphere or the lithosphere (even if it were
deterministic) is very sensitive to the initial conditions. As
these complex systems cannot be fully characterized, they turn out
not to be fully predictable either.

Earthquake prediction \citep{KeilisBorok90,KeilisBL,Rundle}, as
well as some atmospheric predictions \citep{Mason}, is frequently
regarded as a binary forecast: one has to decide whether a large
earthquake is going to occur or not, in a certain time-space
window, instead of calculating the exact probability of this
event. In this binary-forecasting approach, an ``alarm'' is
declared when a large earthquake is expected. If it takes place
when the alarm is on, the outcome is a successful forecast. If it
takes place when the alarm is off, there has been a prediction
failure. If the alarm was declared during a certain period, but
the expected earthquake did not happen, that constitutes a false
alarm.

Note that for using this approach it is necessary to define
precisely what the target earthquakes are that we wish to
forecast. Usually they are defined as those with a magnitude
larger than a given threshold, both when dealing with actual
earthquakes \citep{KeilisBL,Rundle} or with synthetic ones
\citep[e.g.][]{Pepke348,Hainzl}. In the minimalist model, it is
natural to choose as target events the characteristic earthquakes
(size $k=N$), as they mark a distinct peak in the size-frequency
diagram, being much more frequent than other large earthquakes.

A way to quantify the forecasting ability of a certain strategy is
to compute the fraction of errors, $f_e$, and the fraction of
alarm time, $f_a$ \citep{Molchan97}. Given a certain time series
of the model, $f_e$ is the ratio of the total number of prediction
failures to the total number of target events. And $f_a$ is the
ratio of the total time during which the alarm was on to the total
duration of the time series. The fraction of false alarms, $f_f$,
is included in $f_a$, and is the ratio of the total duration of
false alarms to the total duration of the time series.

Of course, a good forecasting strategy should render small $f_a$,
$f_e$ and $f_f$. However, as a general rule, a strategy that
renders low $f_e$ tends to produce large $f_a$ and $f_f$. Dealing
with real seismicity, both a failure and an alarm are costly.
Eventually, decision-makers would need to consider what is less
costly: to predict most of the dangerous earthquakes, but
declaring many alarms, or to declare fewer alarms but failing the
forecast of more large shocks \citep{Molchan97}. Depending on the
trade-off between costs and benefits, one should try to minimize a
loss function, $L$, that can depend on $f_a$, $f_e$ and/or $f_f$.

In the next section, we will describe the forecasting strategies
that will be used to compare the merits of the new strategy
proposed in this paper, based on synchronizing models between
themselves. In the first of the subsections we will indicate the
loss function we will try to minimize in the forecasting of the
model.

\section{Forecasting strategies for comparison}
We describe here three forecasting strategies, based on the
earthquake series, that we will use to asses the merits of the new
strategy described later in this paper. The first two strategies
(the random guessing strategy and the so-called reference
strategy) can be used in any system. The third is specific to the
minimalist model, and serves to ideally determine its maximum
theoretical predictability.

\subsection{Random guessing strategy}

In this strategy, the alarm is randomly turned on and off, during
a certain fraction of alarm time, $f_a$. It is simple to apply
this strategy to any cellular automaton model. Here, in each time
step, the alarm is on with a probability $p$. As a result, the
alarm will be on during a fraction $p$ of time steps ($f_a=p$).
When the target earthquake finally occurs in a certain time step,
there will be a probability $p$ for the alarm to be on. Thus, on
average, a fraction $p$ of target earthquakes will be predicted,
and a fraction $f_e=1-p$ will be prediction failures, so
$f_a+f_e=1$. This strategy has two trivial cases: if the alarm is
always on ($f_a=1$), all the target earthquakes are ``forecasted''
($f_e=0$). Conversely, if the alarm is always off ($f_a=0$), we
fail to predict any of them.

To be statistically significant, any forecasting strategy must
render better results than a random guess. A natural way to
measure this improvement is to consider the loss function
$L=f_a+f_e$. Then, $L=1$ means that the strategy performs as a
random guess, and $L=0$ means a perfect prediction. If $L>1$, the
strategy is performing exactly the opposite to how it should.
Thus, the exact reverse strategy should be considered, and this
will provide the opposite results ($f_a'=1-f_a$, and
$f_e'=1-f_e$).

\subsection{Reference strategy}\label{sec:reference}

Of course, the random guessing strategy depicted above is only
useful as a baseline, but does not serve to provide a real
significant forecast. In this subsection we describe the simplest
meaningful forecasting strategy one can consider for any system.
This will be called the reference strategy, and any forecasting
procedure more complex than this should render better results.

The reference strategy consists simply in declaring an alarm some
time after each target event, and maintaining it on until the next
target event \citep{Newman478,MM2,Box}. As a general rule, the
shorter this time, the bigger $f_a$ and the lesser $f_e$. Which
time is best, then? For the minimalist model, we can look for the
number of time steps $n$ to use with this strategy for obtaining a
smaller $L$. In a previous paper \citep{MM2} we observed that
effectively, for each $N$, there is a $n$ that minimizes $L$. In
Fig. 3, the minimum $L$ that can be obtained with this strategy is
plotted for $N$ between 2 and 20, in the curve labeled
``Reference''. This method does not generate any false alarm, nor
take into account the occurrence of earthquakes smaller than the
characteristic ones. The only information required is the
statistical distribution (probability distribution function) of
the duration of the cycles \citep{MM2}. Taking into account the
effects of smaller earthquakes, the forecast can be modestly
improved in the model \citep{MM2,Europhysics}.

\begin{figure}
\begin{center}
\includegraphics{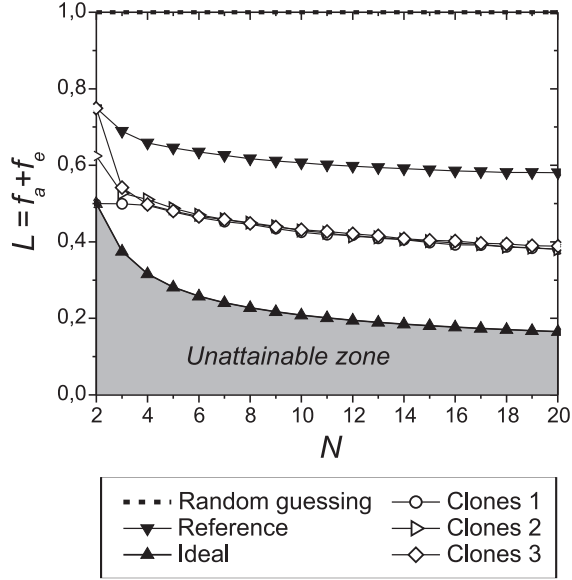}
\end{center}
\caption{Loss function ($L$) obtained with the different
forecasting methods used in this paper, for various system sizes
($2 \le N \le 20$). A random guessing strategy would render $L=1$
for any $N$, while $L=0$ would mean a perfect prediction. The
shadowed zone is unattainable for any forecasting strategy used in
the minimalist model, and the strategy that marks its upper limit
is called ``Ideal''. The ``Reference'' strategy is based only on
the series of the largest earthquakes in the model. The three
strategies labelled ``Clones'' are based on the synchronization of
models with the minimalist model whose largest earthquakes we try
to forecast.}
\end{figure}

\subsection{Ideal strategy}\label{sec:idealstrategy}

As the minimalist model is very simple, it is possible to explore
its maximum predictability. The ideal strategy needed for getting
this result, unlike the two previously described, is
model-specific. It is deduced in Appendix \ref{ideal}. This ideal
result could only be obtained if we could ``see'' inside the model
to check at each time step which cells are occupied and which are
not. Thus it requires a perfect knowledge of the system, and
equivalent strategies cannot be used with actual faults where we
cannot know the detailed state of stress and strain. In Appendix
\ref{ideal} it is deduced that the alarm should be declared at the
instant in which $N-1$ cells of the model are full (just at the
beginning of the plateau with seismic quiescence commented on in
Section \ref{sec:quiescence}). Then, it should be maintained on
until the next characteristic earthquake. This is a no-error
strategy ($f_e=0$, and $L=f_a$). As the model is stochastic, $f_a$
is not zero; a minimum alarm time is needed to forecast all the
characteristic earthquakes. It is given by $f_a=L=N/\langle n
\rangle$, where $\langle n \rangle$ is the average duration of the
cycles (which depends on $N$). This $L$ is also plotted in Fig. 3,
in the curve labelled ``Ideal''. This is the rigorous minimum $L$
that can be obtained in the model. A good forecasting strategy
should produce a $L$ lower than the ``Reference'' curve and as
close as possible to the ``Ideal'' curve.

\section{Synchronization-based forecasting}

In this section we will describe the novel forecasting method
based on the synchronization between models, obtained by imposing
the rupture area of a minimalist model onto other similar models.
This section expands and complements our previous results
\citep{Europhysics}.

We will try to forecast the characteristic earthquakes generated
by a minimalist model with $N$ cells. This model will be called
\textit{master}. We will consider this master as if it were an
actual fault, from which we can know the rupture area of its
earthquakes (equivalent to the number of cells broken, $k$), but
not the strain or stress at depth (equivalent to the occupation
state of the model cells). As in an actual fault, we cannot change
the state of the master at any moment.

In this forecasting method we will use other models, which we call
\textit{clones} \citep{Europhysics}. These are equivalent to the
models that a scientist devises for forecasting the future
evolution of the fault. We will modify their evolution at will,
and their governing rules will be different than those of the
master. In this paper, for simplicity, we will consider that the
clones are also arrays of $N$ cells. The average duration of the
earthquake cycle in the model (average recurrence interval of the
characteristic earthquakes), $\langle n \rangle$, strongly depends
on $N$ \citep{Gomez546}. Choosing a different $N$ for the clones
will imply a different loading rate of the cells and a different
average recurrence interval of the characteristic earthquakes in
the clones than in the master. These effects would require further
tuning of the clones, which would complicate the following
discussion.

Let us describe in the following paragraphs the general outline of
the procedure. We will use a total of $Q$ clones, that will be
loaded (one particle per time step and per clone) at the same time
as the master, but randomly and independently to the master and to
each other. We will apply some procedures for partially
synchronizing the clones with the master. Namely, if in a given
time step the master does not generate any earthquake, we will
oblige the clones not to generate any earthquake either. And if
the master does generate an earthquake, we will force the clones
to reproduce the rupture area of this earthquake, as described
below in more detail. Note that, although the master and the
clones are driven simultaneously, the effects of the master are
dealt with first.

Why use several clones? The master and the clones are all
stochastic, so each one evolves with time in a different way. By
using several clones, we can take into account a broad range of
possible evolutions. By using only one clone, we could not be very
sure that it is satisfactorily mimicking the evolution of the
master. However, if several of these $Q$ clones are in the same
state, then it is more probable that the master is also in that
state. If the clones were deterministic, only one would be
required.

We have commented before (Section \ref{sec:idealstrategy} and
Appendix \ref{ideal}) that the ideal forecasting strategy for the
minimalist model  will be to declare the alarm just when $N-1$
cells of the model become occupied. Then the master enters the
stage of seismic quiescence, or plateau, and the next earthquake
is the characteristic one. We will try to determine this ideal
instant as well as possible with the clones. For this, we will use
a ``democratic'' procedure: we will declare an alarm when a
minimum of $q$ clones ``vote'' (become occupied to a certain
threshold, described below). Later on we will explore the
combinations of $Q$ and $q$ that render the best results. Once the
alarm is declared, it is maintained on until the next earthquake
in the master. If it is a characteristic one, this is a successful
prediction. Its rupture is imposed on the clones (so we reset all
the cells of the clones to empty) and a new cycle starts. If the
next earthquake is not a characteristic one, this represents a
false alarm. We will disconnect the alarm, and impose the rupture
on the clones as is done with any other earthquake. Of course, if
a characteristic earthquake takes place when the clones have still
not declared the alarm (when less than $q$ clones have voted),
this is a prediction failure. If the clones declare an alarm in
the same time step in which the master generates a characteristic
earthquake, we also consider this as a prediction failure.

The exact rules for driving the clones will follow one of the
three approaches commented on below. Each approach imply a
different knowledge of how the master works, and a different way
of imposing the rupture area on the clones. They are depicted in
Fig. 4 and described as follows:
\begin{enumerate}
\begin{item} This first approach will indicate which
is the best result that can be obtained with the
synchronization-based forecasting. For this reason, the clones are
indeed minimalist models identical to the master
\citep{Europhysics}. The clones are loaded only if the master does
not generate an earthquake in that time step. We know that in this
case the particle in the master has gone to one of the cells $i\ge
2$, so the particles in the clones will be randomly thrown to the
cells $i\ge 2$. We also consider as known that, just after an
earthquake with rupture area $k$, the first $k+1$ cells in the
master, for sure, are stable (the $k$ just broken plus the one
that acts as a barrier for the rupture). Thus, if the master
generates an earthquake of size $k$, we will reset to empty the
first $k+1$ cells of the clones. A clone votes when $N-1$ of its
cells are full.
\end{item}
\begin{item} In this second approach, we are more ignorant about
how the master works.  At every time step we will throw the stress
particles to any of the cells in the clones. If the master
generates an earthquake of size $k$, we only know which cells have
ruptured, so we will reset to empty only the first $k$ cells of
the clones. A clone votes when its $N$ cells are full.
\end{item}
\begin{item} In the third approach we know even less. At every
step we will throw the stress particles to any of the cells in the
clones. When an earthquake takes place in the master, we only know
its size, and thus its rupture area, $k$, but not exactly which
cells have ruptured. Thus, we will randomly empty $k$ occupied
cells of each clone. If the clone has less than $k$ occupied
cells, all are emptied. A clone votes when its $N$ cells are full.
In this approach the positions of the cells in the clone are
irrelevant. Each clone is thus equivalent to the so-called box
model \citep{Box}.\end{item}
\end{enumerate}

\begin{figure}
\begin{center}
\includegraphics{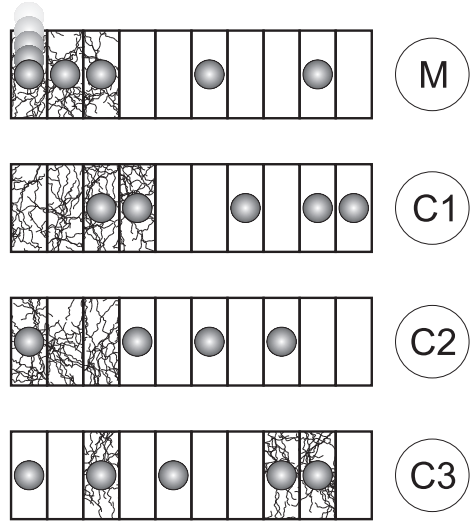}
\end{center}
\caption{Sketch that shows how the rupture area of an earthquake
in the master model (M) is imposed on the clones (C), for each of
the three synchronization-based approaches. In this example, the
master generates an earthquake with rupture area $k=3$. In the
first approach (C1), the first $k+1$ cells rupture and will be
reset to empty. In the second one (C2), this occurs only with the
first $k$ cells. In the third approach (C3), this happens to $k$
occupied cells chosen randomly. The first cell of the clones can
be occupied only in the second and third approaches.}
\end{figure}

Note that, ideally, the clones should have the same number of
occupied cells as the master. For this reason, as a way to measure
the degree of synchronization between a clone and the master, we
used the fraction of time, $\tau$, during which both of them have
the same number of occupied cells \citep{Europhysics}. If two
independent masters run simultaneously, they have the same number
of occupied cells, just by chance, during a certain $\tau$. When a
clone and a master are compared, this $\tau$ greatly increases, as
shown in Fig. 5: partial synchronization is achieved. The best
results, as expected, are achieved with the first of the three
approaches.

\begin{figure}
\begin{center}
\includegraphics{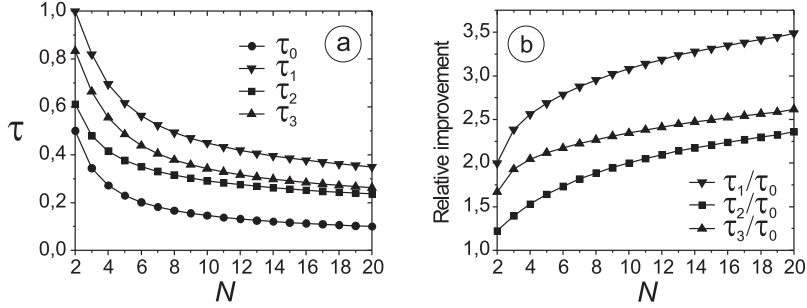}
\end{center}
\caption{{\bf(a)} The fraction of time, $\tau$, during which two
models have the same number of occupied cells. This depends on
whether they are two independent masters ($\tau_0$) or a master
and a clone (governed by one of the three different
synchronization approaches: $\tau_1$, $\tau_2$ and $\tau_3$).
{\bf(b)} The relative improvement, defined as $\tau_n/\tau_0$.}
\end{figure}

The results of $f_a$, $f_e$, $f_f$ and $L=f_a+f_e$, for different
values of $Q$ and $q$ can be plotted as in the diagrams of Fig. 6.
In this figure we have plotted only results corresponding to the
first of the three approaches and $N=20$, but similar figures,
with the same overall properties, can be drawn for the other two
approaches and for any $N$ (see below). There are simple trends in
these graphs. In Section \ref{sec:forecasting} we noted that, in
general, a forecasting strategy that produces lower $f_e$ tends to
produce higher $f_a$ and $f_f$. If $Q$ is fixed (same row), the
greater the $q$, the later the alarm is declared, so $f_a$ and
$f_f$ are lesser and $f_e$ is greater. If $q$ is fixed (same
column), the greater the $Q$, the earlier the alarm is declared,
resulting in the opposite trend.

\begin{figure}
\begin{center}
\includegraphics{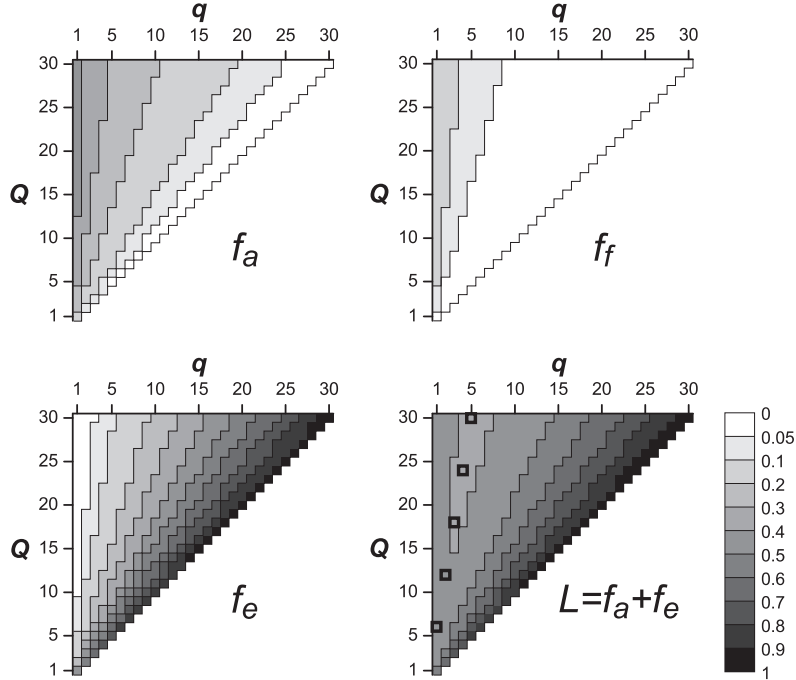}
\end{center}
\caption{Fraction of alarm time ($f_a$), fraction of false alarm
time ($f_f$), fraction of errors ($f_e$), and loss function ($L$)
obtained with the the first synchronization-based forecasting
approach, for $N=20$ and different numbers of clones ($Q$) and
votes ($q$). The squared cells mark a rectilinear valley in the
values of $L$.}
\end{figure}

We are interested in finding the combinations of $Q$ and $q$ that
minimize $L$. The interesting fact is that the sum $f_a+f_e$ shows
a rectilinear ``valley'' for certain combinations of $Q$ and $q$,
marked with squares in the graph of Fig. 6. This valley goes down
as $Q$ and $q$ increase. In Fig. 7 it can be observed that the
valley goes down indefinitely, tending to a lowest asymptotic
value of $L$. We estimate this value, as a function of $Q$, with a
three-parameter exponential fit of the form $F=a\,\exp[b/(Q+c)]$,
where $a$, $b$, and $c$ are parameters. The value of $a$ is the
asymptotic one for $Q\to\infty$. This value is represented, for
each $N$, in Fig. 3. The $f_a$, $f_f$ and $f_e$ also have
asymptotic trends along this valley of $L$, also plotted in Fig 7.
They can also be fitted with the same kind of three-parameter
distribution, to estimate their asymptotic values as $Q\to\infty$.
A nice property is that, as shown in Fig. 7 for a certain case,
these forecasting approaches predict most of the characteristic
earthquakes ($f_e$ is low), and have a very small fraction of
false alarms. Note also that a few tens of clones already render
results close to the asymptotic ones.

\begin{figure}
\begin{center}
\includegraphics{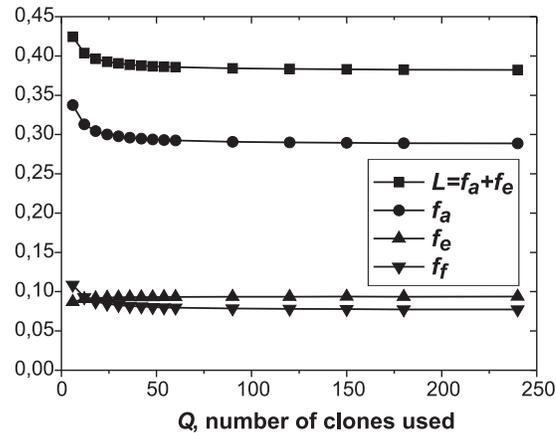}
\end{center}
\caption{Loss function ($L$), fraction of alarm time ($f_a$),
fraction of false alarm time ($f_f$) and fraction of errors
($f_e$) obtained with the first synchronization-based forecasting
approach, for $N=20$ and different numbers of clones ($Q$) along
the rectilinear valley observed in $L$ in Fig. 6 (the first five
points of each curve correspond to the cells marked in that
figure).}
\end{figure}

As can be noted in Fig. 3, the synchronization-based strategies
perform much better than a random guess, and also much better than
the reference strategy described in Section \ref{sec:reference}.
Their results are intermediate between the ideal forecast and the
reference one. The second and third synchronization-based
approaches give only slightly greater $L$ than the first one. The
differences are large only for small $N$. Although the first
approach synchronizes more efficiently each individual clone with
the master (Fig. 5), this effect is compensated by using many
clones.

To assess the performance of the method with larger systems, we
plot in Fig. 8 the results of $L$ for the three approaches, for
$N=100$ and up to 60 clones. As occurred for smaller $N$, a
rectilinear valley is observed in the graphs, and this tendency
can be extrapolated to estimate the asymptotic value of $L$. Note
that the results for the first and second approaches are almost
identical (although $L$ is slightly larger in the second
approach). With the third approach, $L$ decays to its asymptotic
value more slowly (the valley floor has a smaller slope, so more
clones are needed to achieve a given low $L$). The asymptotic
values of $L$, however, are very similar in the three cases (0.298
for the first and second approaches; 0.306 for the third one).
Note that these values are smaller than for $N=20$, as expected
from the trend observed in Fig. 3. The $f_a$, $f_e$ and $f_f$ show
trends similar to the ones described for Figs. 6 and 7. The
asymptotic $fe$ is very low (0.062, 0.075 and 0.071 for the first,
second, and third approach, respectively).

\begin{figure}
\begin{center}
\includegraphics{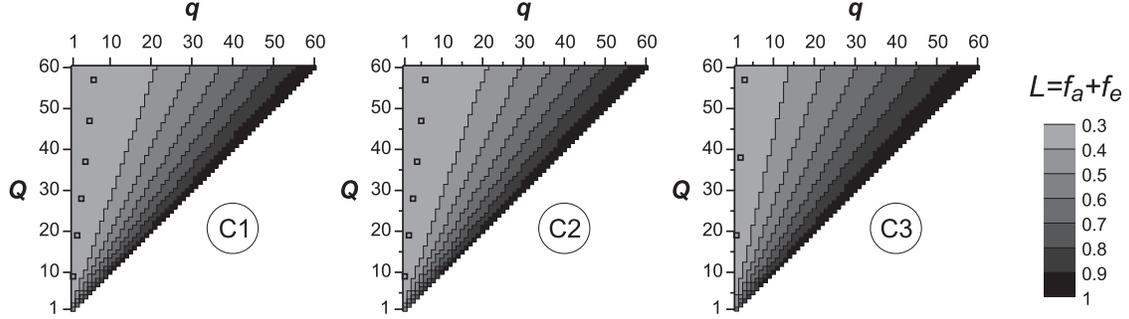}
\end{center}
\caption{Loss function ($L$) obtained with the three
synchronization-based forecasting approaches, for $N=100$ and
different numbers of clones ($Q$) and votes ($q$). The squared
cells mark rectilinear valleys in the values of $L$.}
\end{figure}

Another way to measure the synchronization of the clones with the
master is drawn in Fig. 9 for $N=20$.
The ideal strategy (Section \ref{sec:idealstrategy} and Appendix
\ref{ideal}), would be to declare the alarm just when $N-1$ cells
of the master are full. The figure shows how a single clone
declares the alarm around that moment, but a group of clones does
a much better job.

\begin{figure}
\begin{center}
\includegraphics{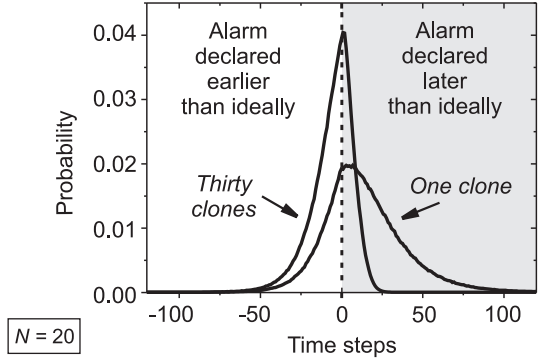}
\end{center}
\caption{Probability that a clone or thirty clones (with $q=5$;
the uppermost squared cell in $L$ in Fig. 6) declare an alarm in a
given time, around the instant when it should for obtaining the
ideal forecast.}
\end{figure}

\section{Discussion and conclusions}

In this paper we have tried to provide some insight into how to
synchronize numerical models with seismic faults, in order to
better forecast large earthquakes in them. The idea is that,
although we can rarely measure the stress and strain in actual
faults, we can estimate the rupture area and coseismic
displacement of their earthquakes. If we force a calibrated model
to reproduce every earthquake rupture of the fault it simulates,
probably the model will be synchronized with the fault. Then it
could be used to forecast the future evolution of the fault,
including future large earthquakes. This idea is not completely
new: e.g. \cite{Ward542} forced a model to reproduce a
large-earthquake rupture and run the model forward to see what
could happen in the future. The results of this paper expand on
earlier ones \citep{Europhysics}, and are still only theoretical,
but fully quantitative. We demonstrate that it is possible to
partially synchronize numerical fault models between themselves,
and use this to forecast synthetic earthquakes.

One of the models, called the master, evolves freely. We consider
it as an actual fault, from which we can know the rupture area of
its earthquakes, but not the strain or stress at depth. Our goal
is to forecast the largest earthquakes it generates. In the
synchronization-based forecasting, we use several other models,
called clones, similar to the master (calibrated to have the same
average recurrence interval of large earthquakes that the master
has). These clones are equivalent to the models that can be
devised to simulate a seismic fault. They are run simultaneously
and independently to the master and to each other. We force them
to reproduce the series of earthquake ruptures of the master, and
this makes them partially synchronized with it. In simple words,
if the master does not generate an earthquake, we preclude any
earthquake in the clones; if the master does generate an
earthquake, we impose the same rupture area on the clones. When
several of the clones indicate that a large earthquake is
impending in the master, we declare an alarm. This efficiently
predicts most of the largest earthquakes of the master, with a
relatively low fraction of total alarm time and few false alarms.
These results are robust: they are almost the same when the exact
rules for imposing the earthquake ruptures vary, and this good
performance is observed along the whole range of model sizes
considered. This synchronization-based forecasting outperforms
other procedures based only on the earthquake series of the model
\citep{MM2,Europhysics}.

The master and the clones are stochastic (random), so each
individual clone is only partially synchronized with the master.
However, when several clones are in the same state, then it is
more likely that the master is also in this state, so the group of
clones makes a much better forecasting job than only one clone
does. If the clones were deterministic, as a general rule only one
would be needed; more clones would have identical evolutions if
run with the same initial conditions.

The procedure developed here is a kind of \textit{ensemble}
forecasting, in which several models are run to obtain a better
picture of how a system will evolve. This concept is used in
atmospheric forecasting  \citep{Palmer615}: several models are run
simultaneously, and their average result has a larger forecasting
ability than that of an individual model
\citep{Houghton,Palmer615}. Each model in this approach has
slightly different initial conditions, to take into account
measurement errors and then to represent one possible state of the
atmosphere, among various possibilities. In our approach, each
clone marks a possible state of the master among a range of
possible options. Several deterministic clones could also be used
with different initial conditions.

Our procedure also shares some similarities with certain
earthquake forecasting algorithms \citep{Kossobokov414,KeilisBL},
in which several seismicity functions are evaluated in real time.
When several of these functions indicate that a large earthquake
is probable, an alarm is declared. In our approach, the clones are
performing a role similar to these functions, monitoring what is
happening in the master.

Our proposal is that a possible way to synchronize more complex,
calibrated models with real faults might be to force them to
reproduce the past series of earthquakes (with the same rupture
area and/or coseismic displacement). This would need to be tested
in the future. Also it will be possible to test whether this
procedure works in the forecasting of synthetic earthquakes in
other models.

Forcing the models to reproduce only one large observed rupture
\citep[as in][]{Ward542} probably is not enough (this is certainly
the case in our stochastic model). Complex and chaotic systems,
such as the lithosphere, are very sensitive to initial conditions.
Forcing the model to reproduce only one rupture is a necessary and
laudable first step, but probably does not constrain the initial
conditions sufficiently. We propose that every observed rupture,
albeit small, should be considered. Small earthquakes are much
more frequent than large ones, thus providing much more data.
Moreover, they provide insight into the mechanical state of the
crust \citep{Seeber595} and into the mechanics of earthquake
rupture \citep{Rubin2002}. Their location may indicate the patch
of the fault plane which is experiencing higher stresses and is
likely to rupture in the next large shock \citep{Schorlemmer588}.
Finally, they are important in the transfer of stress within the
lithosphere, and in earthquake triggering \citep{Helmstetter596}.

\section{Acknowledgments}
\'A.G. thanks Robert Shcherbakov for helping him to view the model
with new eyes. We benefited from reviews by Kristy F. Tiampo and
an anonymous reviewer. Part of the numerical simulations were done
in the computer cluster of the Institute of Biocomputing and
Physics of Complex Systems (BIFI), at the University of Zaragoza.
The Spanish Ministry of Education and Science funded this research
by means of the project FIS2005-06237, and the grant AP2002-1347
held by \'A.G.

\appendix

\section{Deduction of the ideal forecasting
strategy}\label{ideal}

In this appendix we will deduce the ideal strategy outlined in
Section \ref{sec:idealstrategy}. This strategy renders the lowest
(best) value of $L=f_a+f_e$ achievable in the minimalist model.

For this reasoning we would consider every cycle of the model as
composed of two independent and consecutive stages. The first,
that will be called the \textit{loading} stage, starts just after
the occurrence of a characteristic earthquake. During this stage
the total number of occupied cells grows, but not in a monotonic
way, because the particles may land in already occupied cells (and
then be dissipated), and also because of the occurrence of
non-characteristic earthquakes (Fig. 2). When $N-1$ cells (all but
the first one) become occupied, this first stage ends and the
second stage, that will be called the \textit{hitting} stage (or
\textit{plateau} in the occupation), starts. In this second stage,
the system resides statically in the state of maximum occupancy
(Fig. 2) until a particle arrives at the first cell. Then, a
characteristic event occurs, all the cells are emptied, and a new
cycle begins. The hitting stage can be mathematically treated as a
form of Russian roulette.

Both the time spent by the system in the loading stage, $x$, and
in the hitting stage, $y$, are statistically distributed. The
distribution of $y$, denoted by $P_2(y)$, is geometric.
Considering that, in each time step, the probability of hitting
the first cell is $p=1/N$, and its complementary is $q=1-1/N$, it
follows that
\begin{equation}\label{eq1}
P_2(y)=\frac{1}{N}\left(1-\frac{1}{N}\right) ^{y-1},
\end{equation}
whose mean is
\begin{equation}\label{eq2}
\langle y \rangle=N,
\end{equation}
and whose standard deviation is
\begin{equation}\label{eq3}
\sigma=N\sqrt{1-\frac{1}{N}}\,.
\end{equation}

The time elapsed between consecutive characteristic events has
been denoted by $n$, which is statistically distributed according
to the function $P_N(n)$ \citep{VazquezPrada250,MM2,Gomez546}.
Because the variables $x$ and $y$ are independent, the mean length
of the cycles $\langle n \rangle$ is the sum of the mean lengths
of the two stages:
\begin{equation}\label{eq4}
\langle n \rangle=\langle x \rangle + \langle y \rangle.
\end{equation}

It is clear that the best $L$ would be obtained \textit{only if}
we knew the state of occupation of the system and could mark, for
each cycle, the instant at which the stage of loading concludes.
In this case, $f_e=0$, but because the hitting stage is completely
stochastic, $f_a$ (and thus $L$) cannot be nil.

Let us explore the result of $L$ obtained if we turn the alarm on
at a given value $y= y_0$ within the second stage of all the
cycles. With this strategy, the fraction of errors is given by
\begin{equation}\label{eq6}
f_e(y_0)=\sum_{1}^{y_0}P_2(y),
\end{equation}
and inserting Eq. \ref{eq1} we obtain
\begin{equation}\label{eq7}
f_e(y_0)=1-q^{y_0}.
\end{equation}
With respect to to the fraction of alarm, its form is
\begin{equation}\label{eq8}
f_a(y_0)=\frac{\displaystyle\sum_{y_0}^\infty (y-y_0)\cdot
P_2(y)}{\langle x \rangle + \langle y \rangle},
\end{equation}
and inserting Eq. \ref{eq1}, we get
\begin{equation}\label{eq9}
f_a(y_0)=\frac{Nq^{y_0}}{\langle x \rangle+N}\,.
\end{equation}

Note the important contribution of the first stage of the process
in the denominator. Thus, the specific form of the loss function
is
\begin{equation}\label{eq10}
L(y_0)=1-q^{y_0}+\frac{Nq^{y_0}}{\langle x \rangle+N}\,.
\end{equation}
It is noteworthy that in the absence of the first stage, i.e. in
the hypothesis of a pure geometric distribution, the value of $L$
would be 1, not dependent on the value of $y_0$. In this sense,
the geometrical and the Poisson distributions are equivalent. The
minimum value of $L$ in Eq. \ref{eq10} as a function of $y_0$ is
obtained for $y_0=0$, i.e. just after the end of the first stage,
when the $N-1$ upper cells of the system are full. And this
minimum value is
\begin{equation}\label{eq11}
L_{min}=\frac{N}{\langle n \rangle}.
\end{equation}
This result constitutes a rigorous lower bound for the expected
accuracy of any forecasting strategy in the minimalist model. For
this model, $\langle n \rangle$ increases rapidly as $N$ grows
\citep{Gomez546}. This implies that the minimum $L$, obtained with
this optimal forecasting strategy, decreases as $N$ increases, as
shown by the curve labeled as ``Ideal'' in Fig. 3. That is to say,
minimalist models with more cells are more predictable. This is
consistent with the fact that the time series of characteristic
earthquakes is more periodic for larger $N$ \citep{Gomez546}.

\end{document}